\documentclass[]{aa}
\usepackage{graphics,natbib,amssymb}
\usepackage{amsmath}

\begin{document}

\title{An HST study of three very faint GRB host galaxies
\thanks{Based on observations made with the NASA/ESA Hubble Space
Telescope, obtained from the data archive at the Space Telescope
Institute. STScI is operated by the association of Universities for
Research in Astronomy, Inc. under the NASA contract NAS 5-26555.
Based on observations made with ESO Telescopes at the La Silla or
Paranal Observatories under programme ID 60.A-0552.
Based on observations made with the Nordic Optical Telescope, operated
on the island of La Palma jointly by Denmark, Finland, Iceland, Norway
and Sweden, in the Spanish Observatorio del Roque de los Muchachos of
the Instituto de Astrofiscia de Canarias}
}
\author{
A.~O.~Jaunsen\inst{1} 
\and M.~I.~Andersen\inst{2} 
\and J.~Hjorth\inst{3} 
\and J.~P.~U. Fynbo\inst{4}
\and S.~T.~Holland\inst{5} 
\and B.~Thomsen\inst{6} 
\and J.~Gorosabel\inst{7,8,9} 
\and B.~E.~Schaefer\inst{10}
\and G.~Bj{\"o}rnsson\inst{11}
\and P.~Natarajan\inst{12}
\and N.~R.~Tanvir\inst{13}}

\institute{
European Southern Observatory, Casilla 19001, Santiago 19, Chile; {\tt ajaunsen@eso.org}
\and
Division of Astronomy, P.O. Box 3000, FIN--90014 University of Oulo, Finland; {\tt michael.andersen@oulo.fi}
\and
Astronomical Observatory, University of Copenhagen, DK--2100 Copenhagen \O, Denmark; {\tt jens@astro.ku.dk}
\and
European Southern Observatory, Karl-Schwarzschild-Stra{\ss}e 2, D--85748 Garching, Germany; {\tt jfynbo@eso.org}
\and
Department of Physics, University of Notre Dame, Notre Dame, IN 46556--5670, U.S.A.; {\tt sholland@nd.edu}
\and
Institute of Physics and Astronomy, University of Aarhus, DK-8000 {\AA}rhus C, Denmark; {\tt bt@ifa.au.dk}
\and Danish Space Research Institute
  Juliane Maries Vej 30, DK--2100 Copenhagen \O, Denmark;
  {\tt jgu@dsri.dk}
\and
  Instituto de Astrof\'{\i}sica de Andaluc\'{\i}a (IAA-CSIC),
  P.O. Box 03004, E--18080 Granada, Spain;
  {\tt jgu@iaa.es}
\and
  Laboratorio  de  Astrof\'{\i}sica  Espacial  y  F\'{\i}sica
  Fundamental (LAEFF-INTA), P.O.  Box 50727, E--28080, Madrid, Spain;
  {\tt  jgu@laeff.esa.es}
\and
University of Texas, Department of Astronomy, C-1400, Austin, TX--78712, U.S.A.; {\tt schaefer@astro.as.utexas.edu}
\and
Science Institute, Dunhagi 3, University of Iceland IS--107 Reykjavik, Iceland; {\tt gulli@raunvis.hi.is}
\and
Department of Astronomy, Yale University, 265 Whitney Avenue, New Haven, CT 06511, U.S.A.; {\tt priya@astro.yale.edu}
\and
Department of Physical Sciences, University of Hertfordshire, College Lane, Hatfield, Hertfordshire AL10~9AB, U.K.; {\tt nrt@star.herts.ac.uk}
}

\abstract{

As part of the HST/STIS GRB host survey program we present the
detection of three faint gamma-ray burst (GRB) host galaxies based on
an accurate localisation using ground-based data of the optical
afterglows (OAs). A common property of these three hosts is their
extreme faintness. The location at which GRBs occur with respect to
their host galaxies and surrounding environments are robust indicators
of the nature of GRB progenitors. The bursts studied here are among
the four most extreme outliers, in terms of relative distance from the
host center, in the recent comprehensive study of \cite{Bloom2002.1}.
We obtain a revised and much higher probability that the galaxies
identified as hosts indeed are related to the GRBs (P($n_{\rm
chance}$)$=0.78$, following \cite{Bloom2002.1}), thereby strengthening
the conclusion that GRBs are preferentially located in star-forming
regions in their hosts. Apart from being faint, the three hosts
consist of multiple structures, indicative of merging and active
star-formation. For one of the hosts, GRB~980329, we estimate a
photometric redshift of z~$\sim3.5$.

}

\date{\today}
\maketitle


\section{Introduction}
\label{sec:intro}

There are two, possibly fundamentally different, classes of gamma-ray
bursts (GRBs) -- short and long duration bursts (being shorter and
longer than $\sim 1$ second respectively). So far every GRB with an
identified optical afterglow (OA) belongs to the class of long
duration bursts, with the possible exception of GRB 000301C
\citep{Jensen2001.1}. There are two general types of models for
producing GRBs, the first involves the merging of binary compact stars
\citep{Paczynski86.1}, the second is related to the death of very
massive stars \citep{Woosley93.1,MacFadyen99.1,Vietri98.1}. The
observations of SN bumps, Fe K-line and OA localizations close to
star-forming regions favour the latter type of models. Due to the
relatively short lifetime of such massive stars, one expects them to
exist in or very close to star-forming regions where they are born.
Thanks to the extensive observational efforts in pursuing GRB events
at various wavelengths there are now several examples where such
positional correlations between the GRB OA and star forming regions in
the host are seen, for instance \cite{Fynbo2000.1} (GRB~980425),
\cite{Holland1999.1} (GRB~990123), \cite{Hjorth2002.1} (GRB~980613),
\cite{Bloom2001.1}, \cite{Chary2002.1} and \cite{Frail2002.1}
(GRB~010222).  There are, however, a few cases in which the GRB does
not seem to originate from intense star-forming regions (SFRs),
eg. and GRB~990705 \citep{Andersen2002.1}.

In this paper we present localisations and host candidates of three
OAs based on ground data and Hubble Space Telescope (HST) STIS
imaging data from the Cycle 9 program GO-8640 ``A Survey of the Host
Galaxies of Gamma-Ray Bursts'' \citep{Holland2000.1} (data and further
information available at
\texttt{http://www.ifa.au.dk/\~{}hst/grb\_hosts/index.html}).  In
Section~\ref{sec:data} we describe the image processing applied to the
data, in Section~\ref{sec:astrometry} the astrometry resulting in the
OA STIS-image localisations is described. In
Sections~\ref{sec:grb980329} -- \ref{sec:grb990308} we indentify the
hosts, present photometry and discuss the host environment. Finally,
in Section~\ref{sec:summary} we discuss the implications of our
localisations and host identifications. Specifically, we re-compute
using our host identifications, the probability computed in
\cite{Bloom2002.1} that none of the host identifications in that
sample are random galaxies. Finally, we summarize our own results.

\section{Data}
\label{sec:data}

Using the STIS-instrument onboard the HST the GRB-systems were
observed using a four-point dithering pattern with shifts of 2.5
pixels ($\simeq$ 0\farcs127) between exposures.  
The data was pre-processed using the standard STIS pipeline and
combined using the DITHER (v2.0) software \citep{Fruchter2002.1} as
implemented in IRAF\footnote{Image Reduction and Analysis Facility
(IRAF), a software system distributed by the National Optical
Astronomy Observatories (NOAO).}  (v2.11.3) and STSDAS (v2.3). The
STIS images were drizzled using 'pixfrac=0.6' and 'scale=0.5' (giving
a pixel size of 0\farcs0254). Note that drizzling introduces
correlated noise between neighbouring pixels.
All GRB-systems were observed using the STIS 50CCD (hereafter CL)
passband with pivotal wavelength PivW$=5851.5$\AA. GRB 980329 was also
observed using the STIS F28X50LP (hereafter LP) passband,
PivW$=7228.5$ \AA.

The photometry was performed in circular or elliptical apertures, as
appropriate, according to the morphology of the host. The size of the
apertures was selected so as to measure the total flux, by first
choosing a plausible shape and then increasing the size until no gain
in flux could be achieved. The sky was measured in an annulus with
corresponding shape to the object aperture and with inner and outer
annulus 1.5 and 4 times the object aperture.  For the STIS
zero-points, we adopted the values found by \cite{Gardner2000.1} for
the HDF-south.  
The zero-points used were therefore ZP$_{CL}=26.387$ and ZP$_{LP}=25.291$.
Foreground (Galaxy) extinction estimates were computed using the
on-line NED extinction
calculator\footnote{http://nedwww.ipac.caltech.edu/} based on the dust
maps provided by \cite{Schlegel98.1}.

Signal-to-noise (S/N) estimates of the host detections were computed
as the ratio between the measured counts in a circular aperture
centered on the object and the sky variance as measured from 10
similar apertures at random positions (on the sky). The (circular)
aperture diameter used was 19, 19, and 9 for GRB~980329, GRB~980519
and GRB~990308, respectively.  Note that the errors on the photometry
is computed as in IRAF DAOPHOT/APPHOT and does not necessarily
correspond to the S/N estimates.

The host pixel positions were determined by using the
IRAF/APPHOT CENTER task and the 'centroid' algorithm therein.  The
'centroid' algorithm computes the intensity weighted means of the
marginal profiles in {\em x} and {\em y}. The results are given in
Table~\ref{tab:grbpos}.

Throughout this paper we use the following cosmological parameters;
$\Omega_M=0.3$, $\Omega_\Lambda=0.7$, $H_0 = 70$ {km s}$^{-1}$
Mpc$^{-1}$.

\section{Astrometry}
\label{sec:astrometry}

\begin{table*}
\caption{GRB R.A. and Dec. and the best fit pixel localisation (incl. 1-$\sigma$ errors) of the OA in the CL images. For each OA, three reference objects are given including their pixel positions}
\label{tab:grbpos}
\begin{tabular}{l|ll|rlrl|ll|ll}
 & \multicolumn{6}{c|}{OA} & \multicolumn{4}{c}{Host galaxy}\\
Target & R.A. & Dec. & X & & Y & & R.A. & Dec. & X & Y \\\hline
980329 & 07:02:38.02 & +38:50:44.3 & $999.36$ & $\pm 1.34$ & $1032.33$ & $\pm 1.22$ & 07:02:38.07 & +38:50:44.3 & $995.55$ & $1027.35$ \\
ref. 1 & 07:02:38.15 & +38:51:03.7 & $941.32$ & & $1800.06$ & & & & & \\
ref. 2 & 07:02:37.43 & +38:50:34.2 & $1274.41$ & & $635.89$ & & & & & \\
ref. 3 & 07:02:38.97 & +38:50:33.2 & $564.13$ & & $598.21$ & & & & & \\\hline
980519 & 23:22:21.54 & +77:15:43.2 & $1039.19$ & $\pm 0.68$ & $990.77$ & $\pm 0.71$ & 23:22:21.34 & +77:15:43.7 & $1045.27$ & $975.98$ \\
ref. 1 & 23:22:17.78 & +77:15:51.3 & $416.04$ & & $1128.28$ & & & & & \\
ref. 2 & 23:22:20.25 & +77:15:23.8 & $1410.45$ & & $1673.18$ & & & & & \\
ref. 3 & 23:22:27.36 & +77:15:43.7 & $1514.40$ & & $462.34$ & & & & & \\\hline
990308 & 12:23:11.49 & +06:44:04.7 & $813.46$ & $\pm 2.79$ & $922.50$ & $\pm 2.03$ & 12:23:11.49 & +06:44:04.7 & $811.70$ & $921.80$ \\
ref. 1 & 12:23:10.06 & +06:43:50.0 & $1651.271$ & & $341.10$ & & & & & \\
ref. 2 & 12:23:12.30 & +06:44:24.3 & $340.29$ & & $1698.05$ & & & & & \\
ref. 3 & 12:23:10.95 & +06:44:20.8 & $1131.42$ & & $1558.46$ & & & & & \\\hline
\end{tabular}
\end{table*}

The procedure used for {\em relative} astrometry from ground-based
images of the afterglow to the STIS data depends on the available
ground-based data, but for all three objects least squares affine
transformations were used.  An affine transformation is the simplest
possible transformation which allows for deviations from square pixels
in reference and target images.  It is known that some of our
ground-based reference data were obtained with CCDs which have
slightly non-square pixels. As we were not able to establish any clear
correlation between position and astrometric residuals, it is
justified to keep all transformations strictly linear, as is also
preferred with a limited number of tie objects. The accuracy by which
the position of a (point) source in a well sampled image can be
determined is dependent on the signal-to-noise ratio (S/N) with which
the source is detected, and the full width at half maximum (FWHM) of the
point spread function (PSF). For low S/N the astrometric standard
error per axis can be approximated with $\sigma_{\rm pos} =
\sigma_{\rm PSF} / (S/N)$, where $\sigma_{\rm PSF}$ is the standard
deviation of the Gaussian approximating the PSF, which formally equals
FWHM/2.35. For high S/N, accuracy is limited by errors in the detector
pixel geometry, usually at the level of 1/20 of a pixel or less. If
the S/N is significantly above 20 or the FWHM is sampled by less than
about 3 pixels, this approximation of the astrometric error is not
valid. Whenever several individual images of the afterglow is
available, a transformation to the STIS reference frame is established
for each image. As the error in the transformed afterglow position in
the STIS image is always completely dominated by the astrometric error
in the ground-based image, the errors of individual transformed
afterglow positions are in practice independent. The standard
deviation of the transformed positions is therefore an estimate of the
actual astrometric error.

\section{GRB 980329}
\label{sec:grb980329}

GRB~980329 was detected by the {\it BeppoSAX} satellite on 1998 March
29.16 UT.  The radio and optical counterparts were discovered by
\cite{Taylor98.1} and \cite{Djorgovski98.1}, respectively. The latter
claimed the detection of the host galaxy at an apparent magnitude of R
$\sim25.7$. \cite{Palazzi98.1},\cite{Gorosabel99.1},\cite{Reichart99.1}
presented optical and near infrared detections of the OA and found it
to decay at a rate typical for most detected OAs ($\alpha_O \sim
1.2$). This decay slope is in good agreement with that found in
X-rays, where $\alpha_X = 1.35\pm0.03$ \citep{intZand98.1}. The
maximum measured (early) brightness of the OA was I $= 20.8$ and R$=
23.6$, leading to an extremely red colour R$-$I $= 2.8 \pm 0.4$
\citep{Reichart99.1}. Near-infrared (NIR) observations, on the other
hand, showed that the NIR colours are approximately flat. These
measurements led \cite{Fruchter99.2} to argue that the red colour
could be caused by the Ly-$\alpha$ forest if the redshift was
z~$\sim5$. This claim was subsequently challenged by the apparent
non-detection of a Ly-$\alpha$ forest in Keck II spectra
\citep{Djorgovski2001.1}.  Recently, \cite{Yost2002.1} presented
supplementary multi-wavelength broad-band photometry of this burst and
claim to rule out z $\ge5$ based on an afterglow model fit to the
data. The NIR photometry shows that the HST/NICMOS October 1998 data
(GO-7863, PI: A. Fruchter) contains a significant contribution from
the OA and therefore it is not suited for host photometry.  The GRB
has also been found to be heavily extincted by dust
\citep[see][]{Lamb2001.1,Bloom2002.1,Yost2002.1}.

We retrieved NTT/EMMI R-band images of GRB~980329 from the ESO
archive, obtained on March 29.99 and 30.99 \citep{Palazzi98.1}. As the
afterglow was detected at low signal-to-noise in the late-time images,
the astrometry was in this case derived from the combined image. Seven
tie objects were used for the astrometric solution.
The astrometric error, as estimated from the residual of the
tie object fit, is estimated to be about 1.25 drizzled STIS pixels, or
0\farcs03, which should be compared with an expected error of
0\farcs02, as estimated from the S/N of the OA image.  The error in
the transformation from STIS CL to the STIS LP
image is a small fraction of a drizzled STIS pixel and can therefore
be ignored. The best fit localisation in the CL image is given in
Table~\ref{tab:grbpos}.

\begin{figure*}
\centering
\resizebox{\hsize}{!}{
\resizebox{7cm}{!}{
 \rotatebox{0}{
  \includegraphics{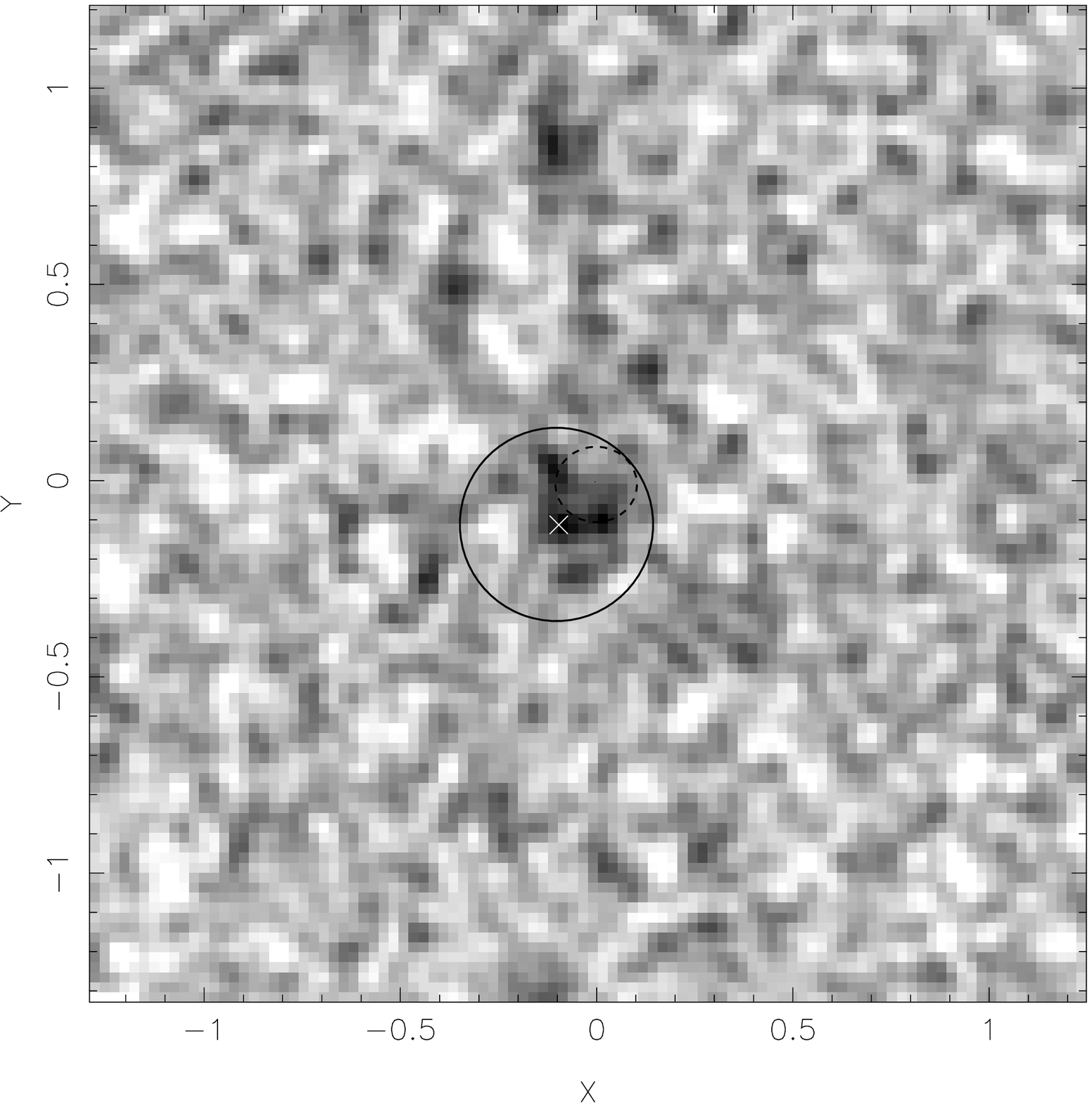}}}
\resizebox{7cm}{!}{
 \rotatebox{0}{
  \includegraphics{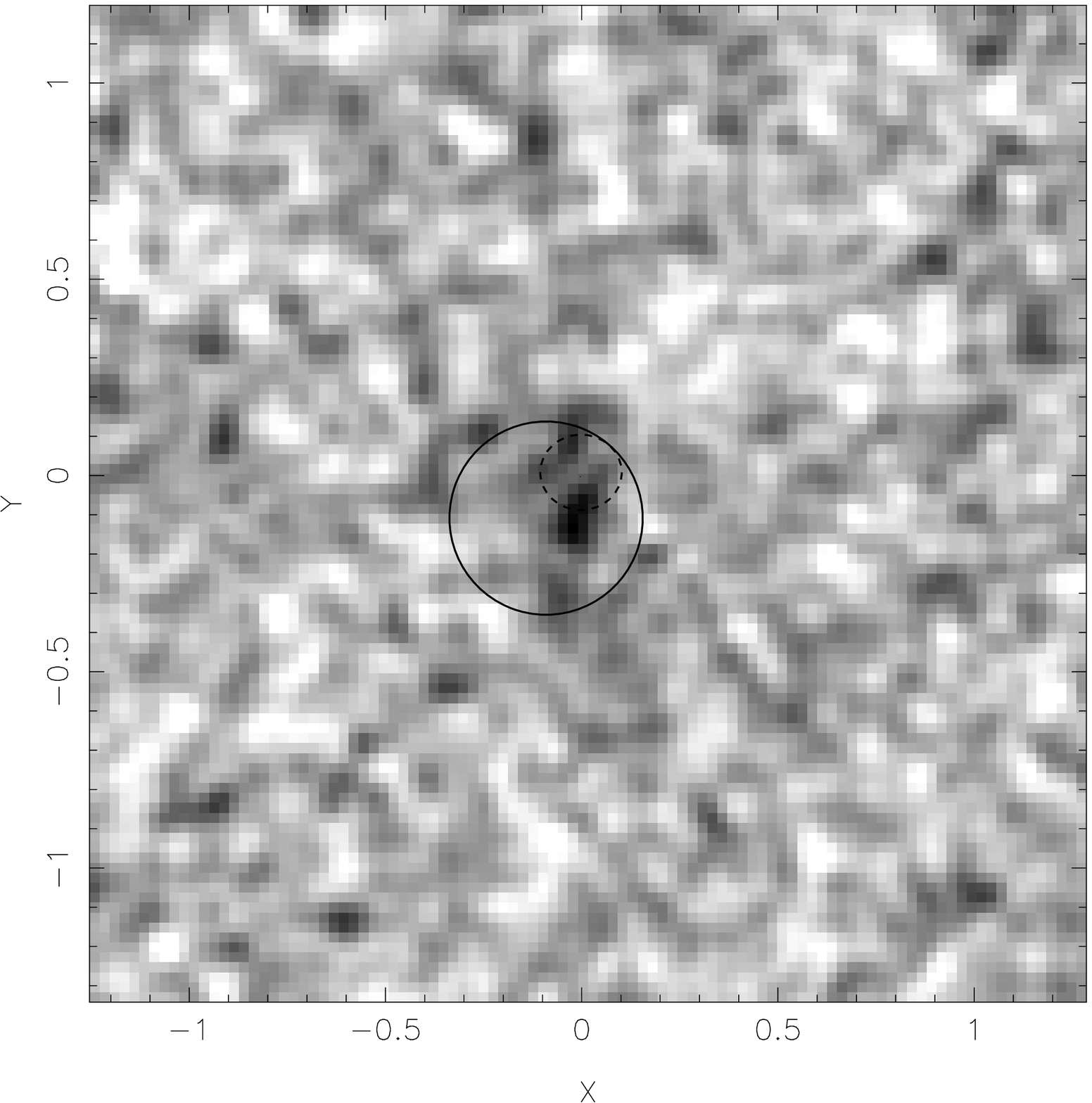}}}
}
\caption{Sub-section of the STIS CL and LP images centered on GRB 980329. A dashed ellipse indicates the 3-$\sigma$ OA localisation error while the solid circle gives the aperture used for photometry and a cross in the CL image marks the host center. The dash-dotted ellipse marks the 3-$\sigma$ error circle of Bloom et~al. All images were smoothed by a 1.1 pixel Gaussian using IRAF/GAUSS. Axis units are in arc-seconds, North is up and East is to the left.}
\label{fig:hosts1}
\end{figure*}

An excerpt of the STIS CL and LP images
are shown in Fig.\ref{fig:hosts1} centered on the host and indicating
the OA position.  In the CL image several unresolved knots are seen on
top of a low surface brightness area within an aperture of
0\farcs5. The measured ABMAG in the CL-band within this aperture is
$27.5\pm0.2$.  In the LP image an extended object is seen, but the
knots seen in the CL-band are not detected. We find $26.6\pm0.2$
within the same aperture in the LP-band.
Foreground extinction corrected photometry and detection signficance
estimates are given in Table~\ref{tab:results}.  Photometry of the
three brightest knots yields a total magnitude of $28.1\pm0.1$ in the
CL-band and $27.7\pm0.2$ in LP. The flux in the LP measurement stems
primarily from the underlying galaxy complex.  Within an arc-second of
the OA position at least two fainter extended objects or structures
are seen to the North and North-East in the CL-band. Their distance
relative to the OA position is approximately 1 and 0.6 arc-second and
photometry measurements yield $28.5\pm0.2$ and $29.5\pm0.5$,
respectively.

The STIS CL and LP measurements in addition to the Keck/ESI R
$=26.53\pm0.22$, I $=26.28\pm0.27$ \citep{Bloom2002.1,Yost2002.1} and
NIRC K $=23.04\pm0.42$ measurements \citep{Yost2002.1} provide an
excellent opportunity to estimate the photometric redshift of the
host.  Using the Bayesian photometric redshift (BPZ) estimation
software of \cite{Benitez2000.1} and restricting z $>1$ \citep[due to
absence of expected emission lines in spectra of the host,
see][]{Yost2002.1} we find z $\sim3.6$ with the best fitting SED
corresponding to an Im galaxy type.  Redshifts of z~$<1.2$ and
z~$>4.2$ are excluded at the 95\% confidence level (z $>5$ excluded at
99.99\% level). These redshift estimates are consistent with the
constraint z~$<3.9$ based on the non-detection of the Ly-$\alpha$
forest in a Keck II spectrum of the host galaxy
\citep{Djorgovski2001.1} and a far-ultraviolet extinction curve
constraint giving $3<$~z~$<5$ \citep{Lamb2001.1}.

It is puzzling that the CL-band shows a clear multi-component nature,
with at least three unresolved knots within 0.5\arcsec, whereas the
LP-band does not. We find an upper limit colour for the knots of
CL-LP~$\le$~0.25, while for the integrated colour of the host complex,
we measure CL-LP~$\approx$~0.8. The compact knots does therefore
appear very blue, implying that Ly-$\alpha$ may be present in the
CL-band. Our best estimate of the redshift is therefore z~$\approx
3.5$. Taking the measured E$_\gamma = 5.5 \times 10^{-5}$ erg cm$^2$
\citep{intZand98.1} we find an isotropic gamma-ray energy of E$_{iso}
= 1.4\times10^{54}$~erg. Assuming a total average energy for GRBs of
$5\times10^{50}$ erg \citep{Frail2001.1} we estimate a jet opening
angle of $\theta_0 = 1.5\deg$, indicating a highly collimated beam.

\section{GRB 980519}
\label{sec:grb980519}

\begin{figure*}
 \centering
 \resizebox{\hsize}{!}{
  \resizebox{5cm}{!}{ \rotatebox{0}{
   \includegraphics{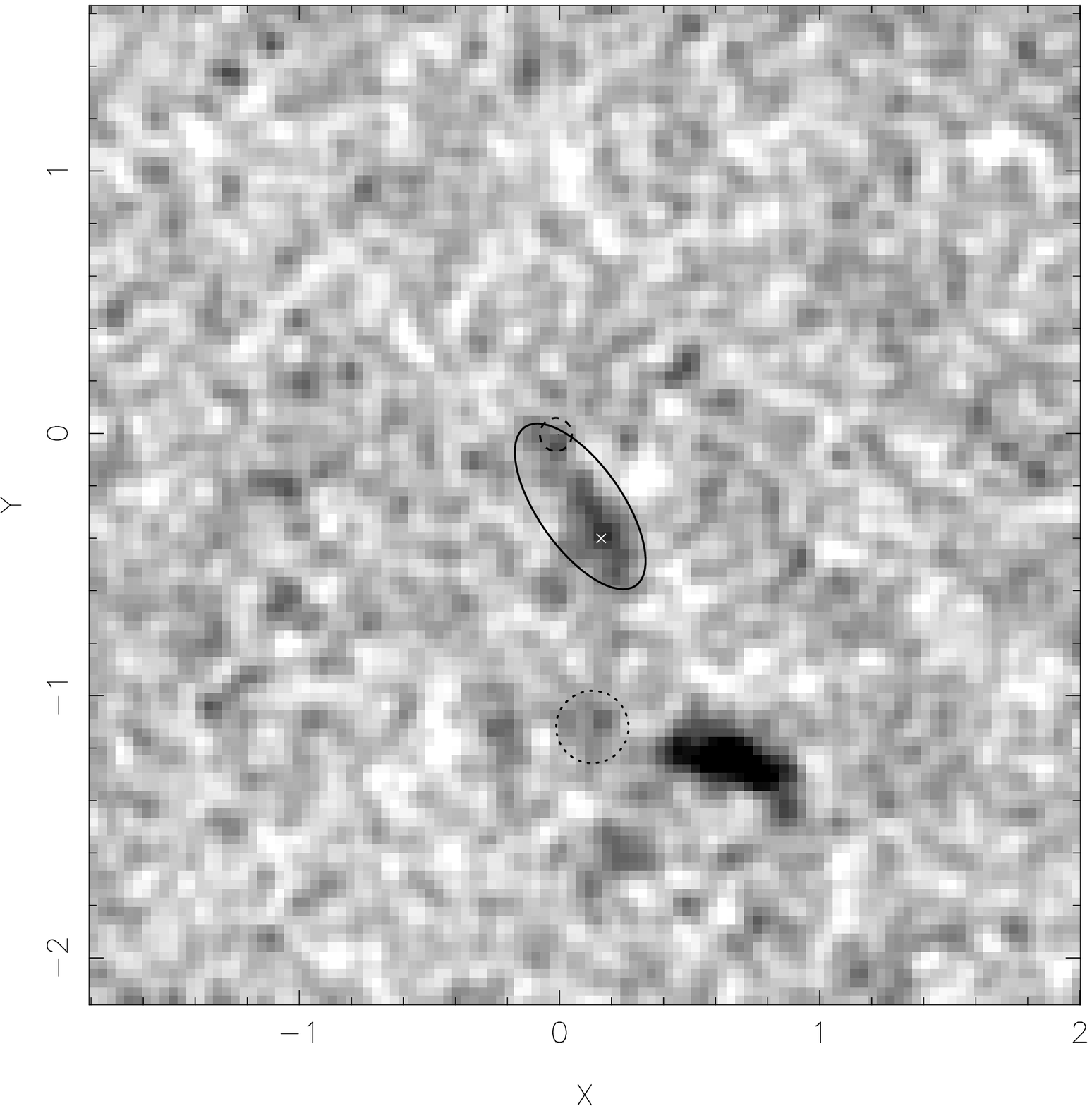}}}
  \resizebox{5cm}{!}{ \rotatebox{0}{
   \includegraphics{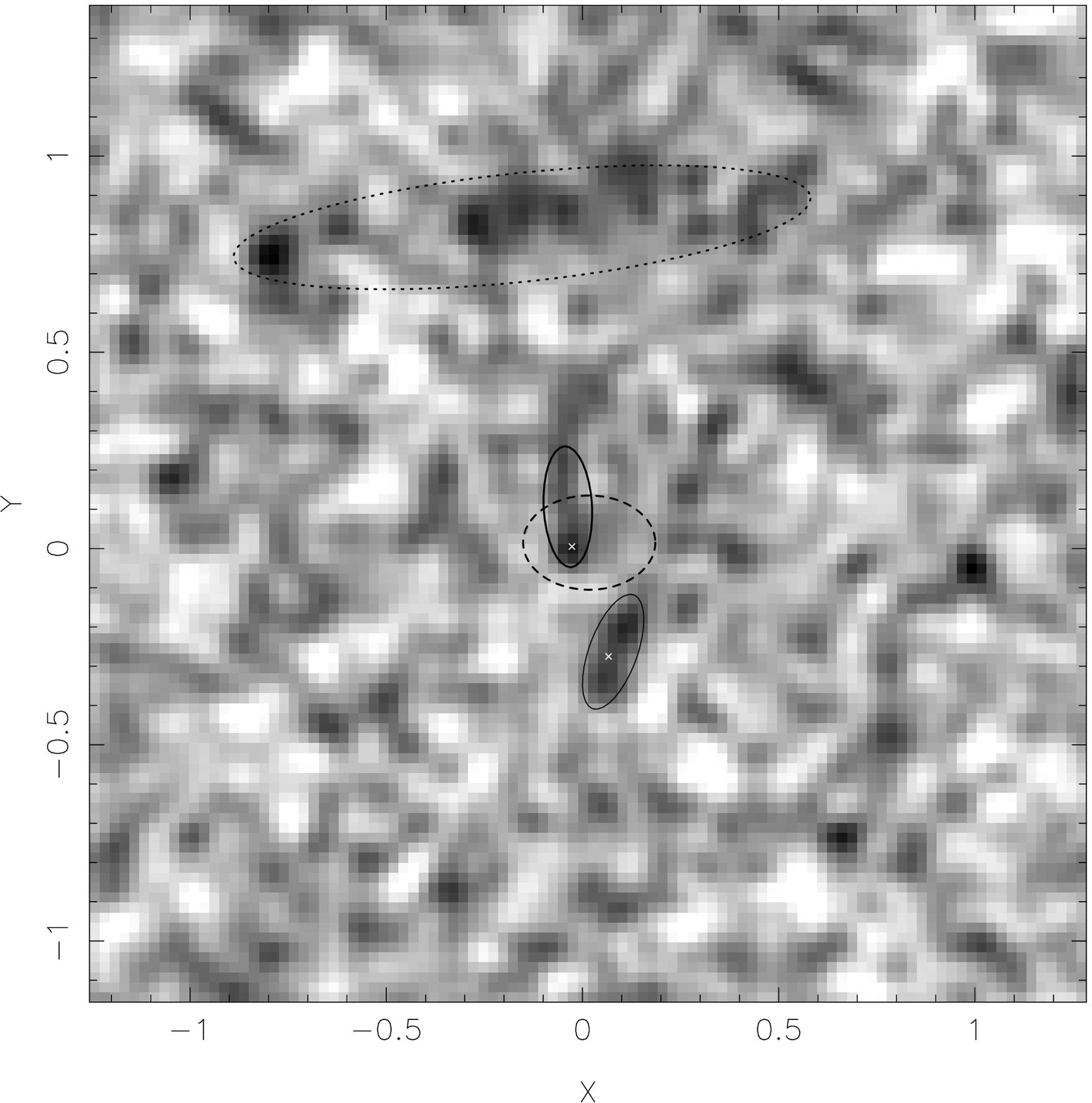}}}
 }
\caption{Sub-section of the STIS CL-band image centered on the OA for GRB~980519 and GRB~990308. Axis units are in arc-seconds, North is up and East is to the left. Annotations as in Fig.~\ref{fig:hosts1} in addition to a dotted ellipse which marks the \cite{Bloom2002.1} host identifications.}
\label{fig:hosts2}
\end{figure*}

GRB~980519 was detected by GRO/BATSE \citep{Muller98.1} on 1998 May
19.51 UT and subsequent {\it BeppoSAX} WFI localisation
\citep{Piro98.1} enabled the detection of the OA \citep{Jaunsen98.1}.
\cite{Jaunsen2001.1} presented a homogeneous optical data sample
obtained at the Nordic Optical Telescope (NOT) and found that the
light-curve breaks at around t$_0 + 0.5$ days, with a pre- and
post-break power law of $\alpha_{\rm O1}=1.73$ and $\alpha_{\rm
O2}=2.22$, respectively.  The X-ray (pre-break) power law was
approximately, $\alpha_{\rm X}=1.6$ though this estimate is somewhat
uncertain as it is based on combining the BeppoSAX WFC and NFI
measurements.  The maximum measured flux of the OA was I$=$18.4 and
R$=$19.8 at t$_0+0.346$ and t$_0+0.536$ days, respectively.  Correcting for
the decline, the approximate R$-$I colour was $\sim 0.6$.

We used the NOT observations of the OA \citep{Jaunsen2001.1} to derive
an accurate position of the burst in the STIS image. As the GRB~980519
afterglow was discovered and followed up at very high airmass,
differential color refraction (DCR) affects the astrometry
significantly. Following \cite{Monet92.1} we correct for DCR by a term
which is proportional to $\sec({\rm Zdist})$ and depends on the color
of each individual object. Since the colors of the tie objects and the
afterglow cover a narrow range, the color dependence was approximated
with a linear relation. The afterglow was observed in the R and the
I bands, so two different DCR corrections must be applied. To minimize
the number of free parameters, we choose to derive a theoretical
estimate of the relative amplitude of the DCR correction in the R and
I bands. From the different refraction across the filter bandwidth it
is found that DCR in R is about 2.65 times DCR in I. By using this
factor, only the amplitude of DCR as function of color for the
complete data set of six images has to be determined.  This is done by
minimizing the standard deviation of the six independent
localisations. For each localisation between 7 and 10 tie objects were
used. The difference between the final averaged positions with and
without DCR correction is of the order of 0.7 drizzled STIS pixels,
comparable to the 1-$\sigma$ astrometric error. In effect, by applying
the DCR correction, the average of the three R-band positions coincide
with the average of the three I-band positions. The OA localisation
result is given in Table~\ref{tab:grbpos}.

At the OA position we detect two extended objects, clearly visible in
Fig.~\ref{fig:hosts2}.  The photometry yields CL ABMAGs of
$27.1\pm0.1$ and $27.9\pm0.1$ for the southern and northern component,
respectively.  The OA is located in the very outskirts of the northern
component (S/N$\sim$9), where a faint blob coinciding with the
position of the OA is detected (see also Table~\ref{tab:results}).  A
host detection was reported by \cite{Sokolov98.1} and \cite{Bloom98.1}
at an estimated Cousins R VEGAMAG of $26.1\pm0.3$. This detection,
however, consisted of the smeared sum of both objects.  By using a
larger aperture of 1\farcs88 (enclosing both objects) we find $\sim
26.45\pm0.10$.  Assuming a flat spectrum and using the STSDAS
SYNPHOT/CALCPHOT we converted the CL ABMAGs to Cousins R VEGAMAG,
giving $\sim 26.0$, in agreement with the earlier combined
detections. It is also worth noting that the combined flux of the two
objects amounts to $26.7$ (ABMAG) as compared to $26.45$ for the large
aperture. The difference in flux can be attributed to the very low
surface brightness ($>29$) regions in the vicinity of the two major
components.  Assuming a redshift larger than 0.5, which is reasonable
given the redshift distribution of other GRB hosts and the faintness
of the host, we note that the average angular scale of 1 arc-second is
$\sim7 \pm 2$ kpc for our assumed cosmology.  Given this angular scale
and the disk-like morphology it is most likely that the two detected
objects are galaxies which are in the process of merging. This is
supported by the low ($<0.003$) integrated probability
\citep{Gardner2000.1} of having two objects of this brightness within
$\sim 2$ sq. arc-seconds.  The faint neighboring patches are therefore
likely to be smaller galaxy fragments belonging to the
merging system.

\section{GRB 990308}
\label{sec:grb990308}

GRB~990308 was detected by GRO/BATSE on 1999 March 8.21 UT, on the
RXTE All-Sky monitor and also weakly by the Ulysses GRB detector. An
OA was detected using the QUEST camera on a 1.0 m Schmidt telescope in
Venezuela \citep{Schaefer99.1}. These optical measurements give a
power law of $\alpha=2\pm2$ \citep{Schaefer99.1}, and are the only
data in which the OA is detected. Early non-detections by LOTIS and
Super-LOTIS suggest $\alpha<1.3$, while later non-detections by the
WIYN and Keck telescopes set the constraint $\alpha>1.1$. Taking all
constraints into consideration, \cite{Schaefer99.1} found a best
fitting constant power-law, $\alpha = 1.2\pm0.1$.

\begin{table*}
\caption{Aperture photometry of GRB hosts in AB magnitudes and corrected for foreground extinction according to \cite{Schlegel98.1} using the Johnson V-band estimate for CL and Cousins I-band estimate for LP.}
\label{tab:results}
\begin{tabular}{l|l|l|l|lll|lll|lll}
GRB & \multicolumn{2}{c|}{Obs. Date} & E(B-V) & \multicolumn{3}{c|}{CL} & \multicolumn{3}{c|}{LP} \\\hline
       & absolute & days & & int. & mag & s/n & int. & mag & s/n \\\hline
980329 & 24/26 Aug 2000 & $\sim 880$ & 0.073 & 8072 & $27.2 \pm 0.1$ & $10$ & 8156 & $26.2 \pm 0.1$ & $9$ \\
980519 & 7 Jun 2000 & $\sim 750$ & 0.240 & 8983 & $27.0 \pm 0.2$ & $9$ & \dots & \dots & \dots \\
990308 & 19 Jun 2000 & $\sim 468$ & 0.023 & 7842 & $29.7 \pm 0.4$ & $5$ & \dots & \dots & \dots \\
\end{tabular}
\end{table*}

We used the original QUEST data \citep{Schaefer99.1} to transform the
OT position to the STIS clear image coordinate system via an
intermediate transformation to a combined NOT image due to the lack of
common tie objects between the QUEST and STIS images. The NOT/ALFOSC
image was combined using standard tools based on 7 R-band images
obtained specifically for this purpose on 29-30 March 2001 with a
total integration time of 6000s.  The final STIS position is the
average of the three individually transformed positions, based on the
three useful QUEST images. The error is estimated from the scatter of
the individual positions and including a contribution from the NOT to
STIS transformation, with the latter being small.  In total 11 tie
objects were used in the transformation to the STIS image. The final
error is consistent with the signal-to-noise of the OT detections in
the QUEST images and the equation introduced in
Sec.~\ref{sec:astrometry}. The final STIS position and error estimate
is given in Table~\ref{tab:grbpos}.  At the locus of the OA we
marginally detect a very faint point-like object (see
Fig.~\ref{fig:hosts2}) which we measure to have a STIS CL magnitude of
$30.1\pm0.4$ with a detection significance of S/N $\sim5$ (foreground
extinction corrected photometry given in
Table~\ref{tab:results}). Including the faint extended emission north
of this object gives $29.9 \pm 0.4$.  We also detect an extended
object 0\farcs3 to the south with an estimated magnitude of
$29.8\pm0.4$ (S/N $\sim3$) and 0\farcs8 to the North a much larger
disk-like object with a magnitude of $27.7\pm0.1$.  Using the galaxy
counts of \cite{Gardner2000.1} we find a relatively low probability
($\sim0.02$ within a radius of 1\arcsec of the OA position) that the
three objects are projected neighbors. 

Could the point-like component coincident with the OA location in fact
be the OA itself?  Assuming a constant power-law one can deduce the
decay slope by interpolating the brightness at the time of the first
V-band QUEST observation and the measured brightness in the STIS
observations 468 days after the burst.  This gives a power-law
exponent of $\alpha \sim 1.35$, consistent with the best estimate of
the power-law slope, $\alpha \simeq 1.2$ (based on all available
data).  If correct, this would be the latest trace to date of an OA,
at t$_0+468$ days.  Another possibility is that the point-like component
is caused by some re-brightening mechanism, such as eg. dust echoing.

In summary, we identify the object coincident with the OA localisation
as the possible remnant OA (point-like) or the host (extended).  If
the point-like object turns out to be non-variable and therefore not
the OA, then it must have been fainter than $\sim30$ mag. This implies
that the late time decay slope must have been larger than 1.35. This
scenario and the constraints from the early data could be explained by
introducing a break in the light-curve. Specifically, an early
$\alpha\sim1.3$ slope (as supported by the early LOTIS data) followed
by a steeper slope fits this scenario well.  A revisit of this field
with HST+ACS is required to disentangle these ambiguities.

\section{Summary}
\label{sec:summary}

We have localised the three OAs to high precision in the STIS images
and identify the host as the nearest detected object of the OA
position.  The GRB~980329 host galaxy redshift is estimated to be
z~$\approx$~3.5.  For all three candidate hosts we detect faint
extended structures within a radius of $\sim 1\arcsec$ ($\sim7$ proper
kpc for z $>0.5$).  This scale is similar to that seen between tidally
interacting and merging galaxies \citep[eg.][]{Borne2000.1}. The hosts
show signs of sub-structure (possibly star-forming and/or merging
elements); 3-4 blue knots in the GRB~980329 host, a knot in the
northern edge of the GRB~980519 host and a point-source within the
3--$\sigma$ localisation error of GRB 990308. The faintness of these
hosts suggests that, regardless of the host luminosity, GRBs seem to
be associated with star formation (SF). This correlation may allow
GRBs to be used as a powerful tracer of star formation, provided the
link between GRBs and SF is correct.  It also implies that a
significant amount of star formation is located in (optically) very
faint galaxies. Their faintness may be due to dust extinction or a
very steep faint-end slope of the galaxy luminosity function. The
SF/GRB correlation may therefore be the basis of a unique and new way
of finding star-forming galaxies at high redshifts (independent of the
otherwise unavoidable surface-brightness bias).

\cite{Bloom2002.1} present a comprehensive study of 20 GRB hosts,
including the three reported here, based on the same HST data and
mainly Keck imaging. For GRB~980329 they found an OA position more
centered on the main host complex compared to our localisation. For
GRB~980519 their OA localisation error is similar to ours but they do
not identify the northern object as the host center. The reason for
this is not entirely clear, but from their thumbnail image (their
Fig. 2) it seems the northern object is much less significant than in
our image and likewise the northern most blob/knot, which is
practically coincident with the OA location.  For GRB 990308
\cite{Bloom2002.1} obtain an OA localisation with significantly larger
errors than ours, and identify a large galaxy in the outskirts of
their error circle (indicated in Fig.~\ref{fig:hosts2}). Our host/OA
localisation corresponds to the object(s) seen left of the center of
the error circle in their image. In summary, we identify different
host candidates in two out of the three faint hosts investigated.
These three hosts are also among the four most extreme outliers (the
ones with the largest OA to host-center offset) in the Bloom et
al. sample. Adopting our host identifications and OA localisations,
the sample does not have obvious outliers.  This is quantified by
re-computing P(n$_{\rm chance}=0$), defined in \cite{Bloom2002.1},
representing the probability that none of the host identifications of
the Bloom et al. sample are randomn galaxies (unrelated to the GRB).
\cite{Bloom2002.1} found P(n$_{\rm chance}=0) = 0.483$, but by using
our OA localisations and host identifications we find P(n$_{\rm
chance}=0) = 0.78$. This means that by using our results it is
unlikely that any of the 20 hosts in the Bloom et~al. sample are false
identifications.

\begin{acknowledgements}
This work was supported by the Danish Natural Science Research Council
(SNF). STH acknowledges support from the NASA LTSA grant
NAG5--9364. JG acknowledges the receipt of a Marie Curie Grant from
the European Commission. MIA acknowledges the Astrophysics group of the
Department of Physical Sciences of University of Oulu for support of
his work.
\end{acknowledgements}

\bibliographystyle{aa}
\bibliography{mnemonic,astro}

\end{document}